\documentclass[twocolumn]{jpsj3}
\usepackage{txfonts}
\usepackage{bm}
\usepackage{color}
\usepackage{booktabs}
\usepackage{comment}

\title{Impact of Ferroelectric Distortion on Thermopower in BaTiO$_{3}$} 

\author{Hiroaki Saijo, Kunihiko Yamauchi\thanks{kunihiko@sanken.osaka-u.ac.jp}, Koun Shirai, Tamio Oguchi}
\inst{ISIR-SANKEN, Osaka University, 8-1 Mihogaoka, Ibaraki, Osaka, 567-0047, Japan}

\newcommand{\bto}{BaTiO$_{3}$}

\abst{
We present a strategy for increasing thermoelectric performance by invoking ferroelectric distortion in a transition-metal oxide. 
By using an {\it ab initio} calculation approach with the Boltzmann transport equation, a large Seebeck coefficient is calculated in $n$-type BaTiO$_{3}$. 
The polar structural distortion causes the peculiar dispersion of the lowest conduction band, which enhances the Seebeck coefficient 
along the polar direction.  
A microscopic mechanism of carrier concentration dependence of the Seebeck coefficient is discussed in terms of the band velocity distribution in the Brillouin zone. 
}

\begin{document}
\maketitle

\section{Introduction}

Due to the recently growing energy issues, thermoelectric power has been attracting much attention 
since it can recover  waste heat and be used as Peltier cooler without producing greenhouse gas emissions. 
As motivated by the advances in the thermoelectric applications, it has been a key issue to improve the thermoelectric efficiency in material-science research field.  
The maximum thermoelectric performance at a temperature $T$ is determined by a dimensionless {\em figure of merit}, $ZT = S^{2}\sigma T/\kappa$, which depends on the Seebeck coefficient ($S$), the electrical conductivity ($\sigma$), and the thermal conductivity ($\kappa$) including the electronic and lattice contributions. 
By assuming a parabolic electronic band and energy-independent scattering approximation\cite{cutler.parabolicapprx}, $S$ is described as 
$S=({8\pi k_{\rm B}^{2}}/{3eh^{2}})m^{*}T(\pi/3n)^{2/3}$, 
where $n$ is the carrier concentration and $m^{*}$ is the effective mass of the carrier.  
The thermal conductivity is further decomposed into the electronic and phonon contributions: $\kappa$=$\kappa_{\rm el}+\kappa_{\rm ph}$.  
Since the $S$ ($\propto n^{-2/3}$) and $\sigma$ ($\propto n$)  show the trade-off dependency on $n$, 
it is not easy to optimize the thermoelectric performance in the range of $n$ limited for the practical use. 
Besides, $m^{*}$ may cause another conflict, 
such that larger effective mass produces higher $S$ but lower $\sigma$. 
Therefore, designing high thermoelectric materials is to find the optimized balance of  $S$, $\sigma$, and $\kappa$.\cite{snyder.review, takabatake.review}  
The conventional strategy for enhancing $ZT$ is to employ heavy  elements and narrow-gap semiconductors as reducing $\kappa$. 
For example, Bi$_{2}$Te$_{3}$ ---now popular as a representative topological insulator\cite{bi2te3.ti.nature}--- shows $ZT \gtrsim$   1 at appropriate temperatures.\cite{tritt.bi2se3} 
These materials are in practical use, however toxic to humans. 

After a striking finding of large Seebeck coefficient in Na$_{x}$CoO$_{2}$\cite{terasaki.nacoo2}, 
transition-metal oxides have emerged as another category of thermoelectrics. 
Electron-doped SrTiO$_{3}$ and KTaO$_{3}$ have been found to have the large Seebeck coefficients $S$\cite{okuda.srtio3, sakai.ktao3}, which are comparable to that of Bi$_{2}$Te$_{3}$, 
as also having the high (metallic) conductivity $\sigma$. 
Several density-functional-theory (DFT) calculations have been performed for the thermoelectric transition-metal oxides, 
as they have mostly concluded that the narrow $t_{2g}$ bands with the large $m^{*}$ are responsible for the large Seebeck coefficient.\cite{singh.srrh2o4, singh.gete}  
Later on, Kuroki {\em et al.} have pointed out that a particular {\em pudding-mold} $a_{1g}$ band may enhance both $S$ and $\sigma$ in dumbbell-type structure in Na$_{x}$CoO$_{2}$ and CuAlO$_{2}$.\cite{kuroki.nacoo2.jpsj, kuroki.cualo2}  
As for SrTiO$_{3}$ and KTaO$_{3}$, they  claimed that the multiplicity of three-fold $t_{2g}$ state plays a role in enhancement of $S$ and $\sigma$. 
When the electron is doped into the empty $t_{2g}$ bands, the multiple bands can keep low Fermi level, which in turn results in a large Seebeck coefficient, while the doped electrons increase the conductivity.\cite{kuroki.srtio3}
Recently, Shirai {\em et al.} casted another interpretation on the large thermopower in SrTiO$_{3}$ as following.\cite{shirai.srtio3}   
It is often said that the $t_{2g}$ components of the $d$ electrons form dispersion-less and three-fold degenerate bands. However, the lift of the degeneracy as leaving from the $\Gamma$ point should not be overlooked. As far as the electron doping is concerned, only the lowest band is relevant. More importantly, even a single band can have electrons of quite different masses in different regions of the Brillouin zone.
The light electron leads to large $\sigma$, whereas the heavy mass  results in large $S$,  
so as to increase the power factor $S^{2}\sigma$. 
If this idea works, it is more critical to have strongly anisotropic $m^{*}$ in the single band rather than multiply degenerate bands for enhancement of the thermopower in oxides. 
In this context, we focus on the $t_{2g}$ bandstructure  in  BaTiO$_{3}$, 
in which the ferroelectric crystal distortion is expected to enhance anisotropy of the $m^{*}$   and to enhance the thermoelectric properties. 
In fact, an early experimental finding of the anisotropic Seebeck coefficient in BaTiO$_{3}$ was reported in 1967\cite{berglund.bto.seebeck}, 
whereas the accurate theoretical approaches to understand the electronic property had not been available yet in that time. 
In the present study, we aim to confirm the effect of the polar distortion on the $t_{2g}$ band and the resulting thermoelectric properties in BaTiO$_{3}$ and 
then we discuss the microscopic mechanism of enhancement of $S$ and $\sigma$  by means of first-principles DFT calculations. 

\section{ Bandstructure and Wannier Interpolation}

\begin{figure}[h]
\begin{center}
\includegraphics[width=0.9\columnwidth, angle=0]{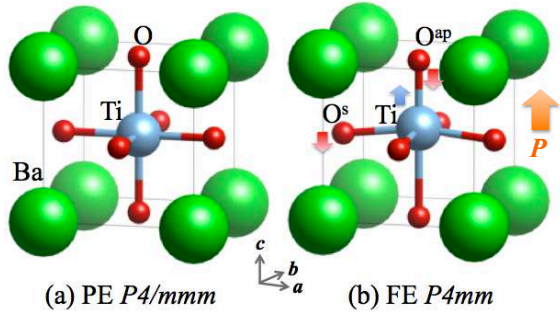}
\end{center}
\caption{
Tetragonal unit cell of BaTiO$_{3}$ in  the (a) paraelectric (PE) structure and (b) ferroelectric (FE) structure. 
Apical and side oxygens  are denoted as O$^{\rm ap}$ and O$^{\rm s}$, respectively. 
Directions of the polar atomic distortions and the polarization are shown by block arrows. The polar displacements are exaggerated by the factor of two for clarity. }
\label{fig:cryst}
\end{figure}

\begin{figure}[h]
\begin{center}
\includegraphics[width=0.8\columnwidth, angle=0]{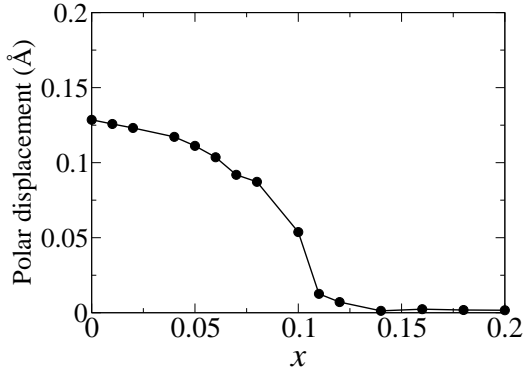}
\end{center}
\vspace{0.2cm}
\caption{
Polar ionic displacement, defined as a change in  bond length between Ti and O$^{\rm ap}$ ions in BaTiO$_{3}$,  
  as a function of the electron doping rate $x$. 
}
\label{fig:FEvsDoping}
\end{figure}

Perovskite \bto\ is a popular ferroelectric oxide, which undergoes a ferroelectric transition at 393K.  
To obtain the electronic structure of ferroelectric \bto, we performed DFT calculations by using the {\tt VASP} code\cite{vasp} with the generalized gradient approximation Perdew-Burke-Ernzerhof (GGA-PBE) potential.\cite{pbe}  
The tetragonal lattice constants $a$=3.991\AA\ and $c$=4.035\AA\ were taken from experimental result at room temperature.\cite{cheong.batio3} 
The internal atomic coordinates were optimized until forces acting on atoms were less than 1$\times10^{-3}$ eV/\AA\ 
both for the paraelectric (centrosymmetric) structure and ferroelectric (polar) structure as shown in Fig.\ref{fig:cryst}.  
A (12, 12, 12) $\bm k$-point mesh was used for the Brillouin zone integration. 
Although the real crystal structure at the paraelectric phase ($T>393$K) is known as cubic, 
we used the same tetragonal lattice for both paraelectric and ferroelectric structures to discuss the effect purely caused by the polar ionic displacement. 
The ferroelectric polarization in non-doped BaTiO$_{3}$ was calculated by Berry phase approach\cite{berry} as 
 $P^{\rm DFT}$=26$\mu$C/cm$^{2}$ along the $c$ axis, which is in good agreement with the experimental value of $P^{\rm exp}$=27$\mu$C/cm$^{2}$\cite{wieder.batio3} 
 and  the previously calculated values of  $P^{\rm DFT}$=22-29$\mu$C/cm$^{2}$\cite{fechner.batio3.P, wang.batio3.P}. 
The polarization can be decomposed into the ionic and electronic contributions:   
 $P^{\rm DFT}=P^{\rm ion}+P^{\rm elec}$, where the ionic contribution was calculated by a point-charge model 
 as assuming the nominal ionic valence as +2, +4, -2 for Ba, Ti, O ions, respectively. 
 The calculated $P^{\rm ion}$=14 $\mu$C/cm$^{2}$,  which is almost half of the total polarization, originates from the Ti and O ionic displacement due to the strong hybridization between occupied O-$p$ and unoccupied Ti-$d^{0}$ state as called {\em $d^{0}$-ness} mechanism.\cite{nicola.d0ness}   
Figure \ref{fig:FEvsDoping} shows the polar ionic displacement as a function of electron doping. 
When electron is doped in BaTiO$_{3}$, the Fermi energy moves in the unoccupied Ti-$t_{2g}$ bands. 
It screens the electric potential of ionic charge and spoils the ferroelectricity. 
Nevertheless, the polar ionic displacement (i.e. the ionic contribution to $P$) is rather robust so as to  stay almost 90\% even when 0.05 $e$ of electron is doped in the Ti-$d^{0}$ state, as consistent with a previous theoretical study in Ref.\citen{tsymbal.dopedbatio3}. 
This result ensures that the crystal structure keeps the polar distortion under some amount of electron doping (e.g. $x\lesssim0.1$), 
which is sufficient for the following discussions on the electron-doped BaTiO$_{3}$ in the ferroelectric structure. 
Technically speaking, 
the cubic-tetragonal (centrosymmetric-polar) transition temperature may decrease by electron doping 
although the polar ionic distortion will persist below the transition  temperature. 
It was reported that the transition temperature is kept over room temperature until 0.022 $e$/u.c. doping, 
while higher doping may lead to lower transition temperature.\cite{kolodiazhnyi.dopedbatio3}  
For the thermoelectric application of BaTiO$_{3}$, tuning of the polar property under electron doping will be required. 
For example, both the polar distortion and  the transition temperature can be largely increased by epitaxial strain in BaTiO$_{3}$ thin film.\cite{miyazawa.batio3.strain, damodaran.batio3.strain} 

\begin{figure}[h]
\begin{center}
\includegraphics[width=0.9\columnwidth, angle=0]{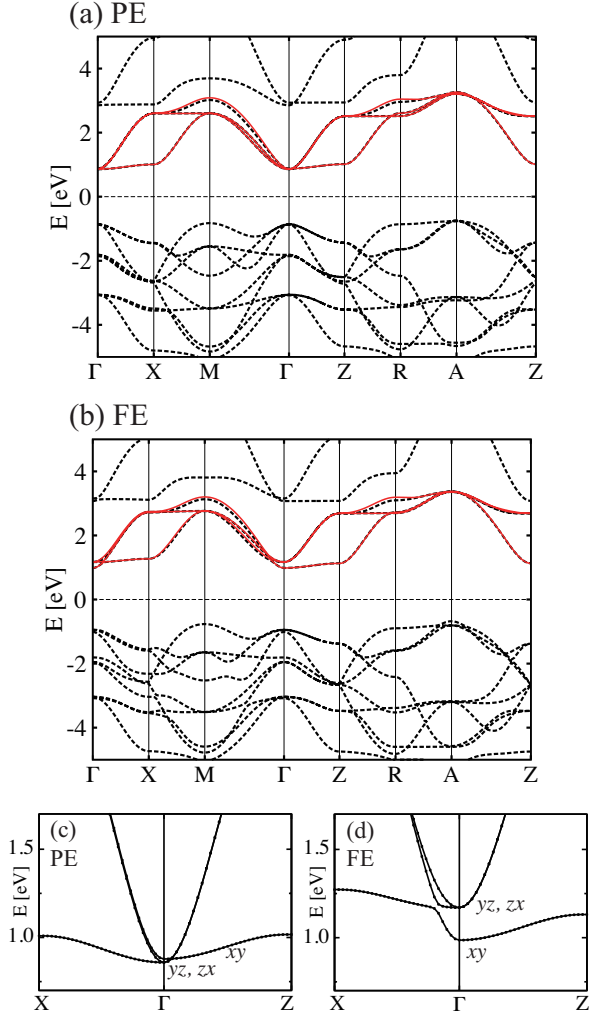}
\caption{
Bandstructure of BaTiO$_{3}$ in (a) paraelectric and (b) ferroelectric phase.  (c) and (d) are their blow-up versions.  
The Wannier-interpolated Ti-$t_{2g}^{}$ bands are superposed on the DFT bandstructure, as highlighted by (red) solid lines in (a) and (b). 
}
\label{fig:band}
\end{center}
\end{figure}

Figure \ref{fig:band} shows the bandstructure of \bto\ in the paraelectric and ferroelectric structure. 
They consists of nine O-p bands below the Fermi energy, three Ti-$t_{2g}$ and two $e_{g}$ bands above the Fermi energy. 
Three Ti-$t_{2g}$ bands lying at the conduction band bottom are well isolated from the other bands. 
Due to the tetragonal symmetry of the crystal structure, the $t_{2g}$  bands  are split into ($yz$, $zx$)  and  $xy$ bands at the $\Gamma$ point. 
The former can be further split  along the $\Gamma$-$\rm X$ line, where the four-fold rotation symmetry is  broken. 
The  symmetry-lowering effect is significantly enhanced in the ferroelectric structure.  
It is intriguing that the ($yz$, $zx$) bands are the conduction bottom in the paraelectric structure, 
whereas  the $xy$ band comes to the bottom in the ferroelectric structure. 
This is caused by strong Ti-O$^{\rm ap}$ hybridization, which pushes up the anti-bonding ($yz$, $zx$)  bands in the ferroelectric structure. 
In Fig.\ref{fig:band} (d), it is observed that the $xy$ and $zx$ bands (which belong to the same irreducible representation $\Delta_{2}$)  repel each other along the $\Gamma$-$\rm X$ line as resulting in the particularly non-harmonic band shape. 
The lowest $t_{2g}$ band is responsible for the thermoelectric properties since the electron will be doped into this band. 
For the further analysis of the thermoelectric property, the $t_{2g}$ bands were interpolated by means of 
maximally localized Wannier functions (MLWF) by using the {\tt Wannier90} code.\cite{vanderbilt.wannier, wannier90}  

\begin{figure} 
\begin{center}
\includegraphics[width=0.70\columnwidth, angle=0]{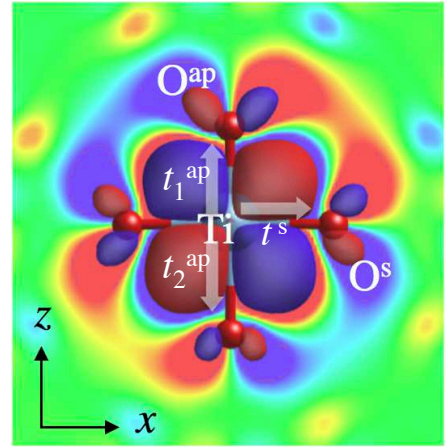}
\caption{
An isosurface of MLWF for Ti- $d_{zx}$ orbital state hybridizing with the surrounding O-$p$ states in the ferroelectric structure. 
The hopping integral between the $d_{zx}$ and the side (s) and apical (ap) oxygen $p$ states are shown by arrows. 
}
\label{fig:wavef_btofe}
\end{center}
\end{figure}
\begin{center}
\begin{table}
\large
\begin{tabular}{|c|rrr|rrr|} \hline
	&   $t^{\rm s}$ & $t_{1}^{\rm ap}$ & $t_{2}^{\rm ap}$ & $l^{\rm s}$ & $l_{1}^{\rm ap}$ & $l_{2}^{\rm ap}$  \\  \hline
	   PE & 1.03 & 0.98 & 0.98 & 2.00 & 2.02 & 2.02 \\
	   FE & 1.03 &  1.34 & 0.72 & 2.00 & 1.89 & 2.15   \\ \hline
\end{tabular}
\caption{Calculated hopping integrals $t$ (eV) and the bond length $l$ (\AA) between Ti-$d_{(yz, zx)}$ and side (s) and apical (ap) O-$p$ orbital states in the paraelectric (PE) and ferroelectric (FE) structures. }
\label{tbl:hopping}
\end{table}
\end{center}
Figure \ref{fig:wavef_btofe} shows one example of calculated MLWF for  Ti-$d_{zx}$  state. 
As related, Table \ref{tbl:hopping} shows the hopping integral between Ti-$d_{zx}$ and O-$p$ states as calculated via the off-diagonal Hamiltonian element for the MLWF basis together with the Ti-O  bond length. 
In the ferroelectric structure, a Ti ion is slightly displaced toward one of the apical oxygens by 0.13 \AA, which considerably enhances the hopping integral. 
Therefore, the MLWF of Ti-$d_{zx}$ has larger weight  at  the top apical oxygen site than the bottom apical oxygen site. 
As shown in Table \ref{tbl:hopping}, the calculated hopping with closer O-$p$ state is almost double of the other ($t^{\rm ap}_{1}>>t^{\rm ap}_{2}$). 
The anisotropic enhancement of $t$ is directly related to the $d^{0}$-ness microscopic origin of ferroelectric distortion, by which the energy gap is increased to lower the band energy and to stabilize the ferroelectric structure.\cite{nicola.d0ness}  
The anisotropic hopping integral slightly modulates the bandstructure but has a great impact on the thermoelectric property. 

\section{Thermoelectric Property}

The standard Boltzmann's transport  approach was adopted in the present study to calculate the Seebeck coefficients.\cite{singh.boltzman} 
In this approach, the Seebeck coefficient is given by 
\begin{equation}
    {\bm S} (\mu , T) =\frac{1}{eT}\frac{{\bm K_{1}}(\mu , T)}{{\bm K_{0}} (\mu , T)}, 
\end{equation}
where $e$ is  electron charge, $T$ is  temperature, $\bm K_{0}$ and $\bm K_{1}$ are given by 
\begin{equation}
      {\bm K}_{i}(\mu, T)=\int_{-\infty}^{\infty} d\varepsilon \sum_{n\bm k} { \tau {\bm v}_{n\bm k} {\bm v}_{n\bm k} \left[ - \frac{\partial f(\varepsilon, \mu,T)}{\partial \varepsilon} \right] (\varepsilon _{n\bm k}-\mu)^{i} } 
\end{equation}
for $i=0,1$. 
Here, $\varepsilon_{n\bm k}$ is the band energy, ${\bm v}_{n\bm k}$ is the band velocity, $\tau$ is the quasiparticle lifetime assumed as constant, $f(\varepsilon) $ is the Fermi distribution function, and $\mu$ is the chemical potential. 
In order to calculate the accurate band velocity (with band index $n$) 
\begin{equation} 
   {\bm v}_{n\bm k}=\frac{1}{\hbar}\frac{\partial \varepsilon_{n\bm k}}{\partial \bm k},  
\end{equation} 
Wannier function approach implemented in  {\tt BoltzWann}\cite{boltzwann} code  was employed.  
The  $\bm k$-point mesh was increased up to  (160, 160, 160) to reach the convergence. 
The electrical conductivity $\bm \sigma$ is also calculated as 
\begin{equation}
 {\bm \sigma}  (\mu , T)= \frac{e^{2}\tau}{V} \int_{-\infty}^{\infty}  d\varepsilon(-\frac{\partial f(\varepsilon,\mu,T)}{\partial \varepsilon} )           \sum_{n,{\bf k}}  {\bm v}_{n\bm k}^{2}  \delta(\varepsilon - \varepsilon_{n,{\bm k}}).  
\end{equation}

\begin{figure}[h]
\begin{center}
\includegraphics[width=0.80\columnwidth, angle=0]{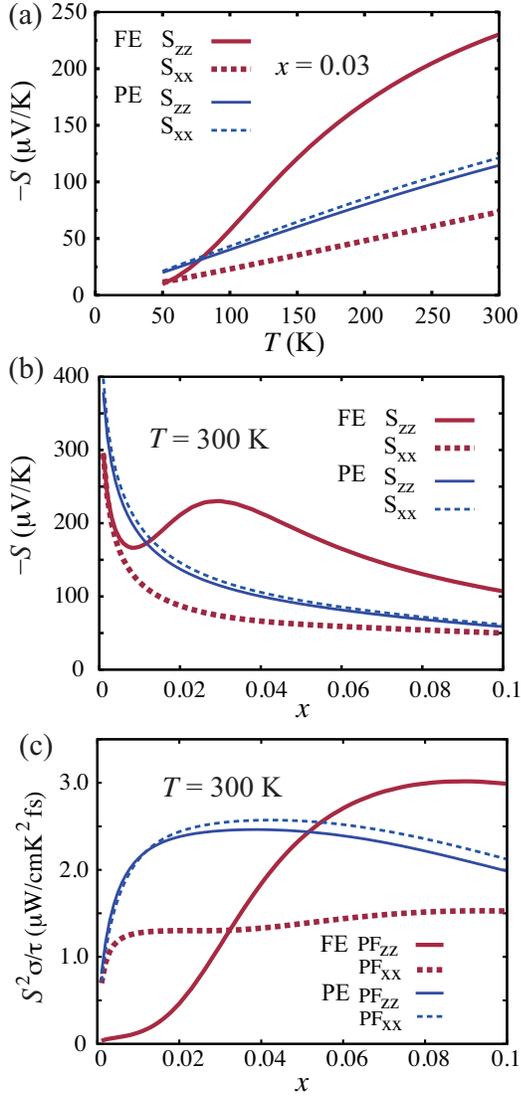}
\caption{
(a) Calculated Seebeck coefficient for  $x$=0.03  plotted as a function of temperature for ferroelectric (FE) and paraelectric (PE) crystal structures.  
 (b) Calculated Seebeck coefficient and (c)  power factor (PF) (devided by $\tau$) as a function of the doping rate $x$ at $T$=300 K. 
 The diagonal matrix elements $zz$ (the polar axis) and $xx$ are plotted by solid and dotted lines, where 
 broad and narrow lines correspond to FE and PE structures, respectively. 
}
\label{fig:seebeck:T}
\end{center}
\end{figure}

In Fig.\ref{fig:seebeck:T} (a), we show the calculated Seebeck coefficients 
as a function of temperature $T$ at a fixed doping rate $x$=0.03. 
In this calculation, the rigid band approximation was assumed. 
In the paraelectric structure,  $S_{xx}$ and  $S_{zz}$ show the similar values. 
$S_{xx}$  reaches $-$122 $\mu$V/K at $T$=300K and $x$=0.03, 
being consistent with  the  previously calculated value for SrTiO$_{3}$ as $S_{xx}$  = $-$87$\mu$V/K at $T$=300K and $x$=0.05.\cite{kuroki.srtio3} 
However, as it was discussed in Ref.\citen{kuroki.srtio3}, the calculated $S$ value for  SrTiO$_{3}$  is almost half-reduced from the experimental result. 
This is  due to the strong correlation effect of the 3$d$ state, which band width is usually overestimated in DFT calculations. 
As for the further correspondence to the real value of $S$, we should note that the $S$ value predicted by DFT calculations is underestimated roughly by the factor of two. 
For the ferroelectric case, the calculated Seebeck coefficient shows the remarkable anisotropy, $S_{zz} \gg S_{xx}$. 
At $T$=300K and $x$=0.03, $S_{zz}$  is calculated as -230$\mu$V/K, which is almost four times larger than  $S_{xx}$ = -74 $\mu$V/K. 
This value may be comparable to the experimentally measured $S\approx$300$\mu$V/K in epitaxial La-doped BaTiO$_{3}$.\cite{gilbert.epi.bto.seebeck}  
Figure \ref{fig:seebeck:T} (b) shows the calculated Seebeck coefficient as a function of the doping rate $x$. 
The $S_{zz}$ coefficient behaves anomalously as increasing $x$, as making a large ``bump'' at $x$=0.03. 
This is caused by the anisotropic property of the Fermi surfaces and the effective mass due to the ferroelectric ionic displacement, as explained in later section. 
It results in a remarkable crossover of power factor ($S^{2} \sigma$) between the paraelectric and ferroelectric phases at $x\sim$0.05 as shown in Fig. \ref{fig:seebeck:T} (c).

\section{Effect of Ferroelectric Distortion}

\begin{figure}[h]
\begin{center}
\includegraphics[width=0.98\columnwidth, angle=0]{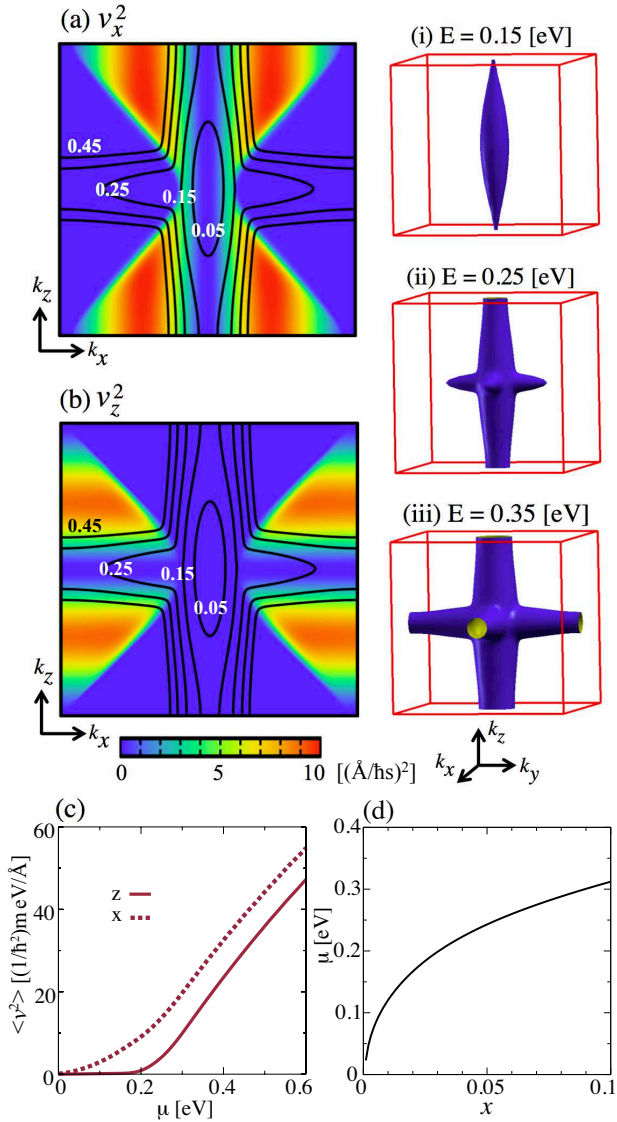}
\caption{
Contour plot of $\varepsilon(\bm k)$=$\mu$ surfaces with several $\mu$ values (origin of the chemical potential $\mu$ is taken as the bottom of the conduction bands) and the squared band velocity plotted by color for (a) $v_{x}^{2}$ and (b) $v_{z}^{2}$ of the lowest conduction band ($xy$ band) in ferroelectric \bto. 
The three-dimensional energy surfaces are shown in the right panel (i)-(iii). 
(c) 
 Squared band velocity $\langle v_{x}^{2} \rangle$ and  $\langle v_{z}^{2} \rangle$ integrated in the $k$-space as 
  $\langle v_{ij} \rangle ^{2} (\mu) = (1/V)   \sum_{n,{\bf k}}  { v}_{in{\bm k}}{ v}_{jn{\bm k}} \delta(\mu - \varepsilon_{n{\bm k}}) $ 
 for the lowest conduction band. 
 (d) Chemical potential  $\mu$ as a function of the doping rate $x$ at $T$=300K. 
}
\label{fig:fermi:v}
\end{center}
\end{figure}

Here, we discuss the microscopic origin of the enhancement and  the anisotropic behavior of Seebeck coefficient $\bm S$ in ferroelectric \bto. 
We note that the enhancement of $S_{zz}$  originates from a single-band contribution.  
In the energy region of $0<\mu < 0.5$ eV, a main portion of the electron density of states originates from the $xy$ band at the conduction bottom, whereas  a contribution from the ($yz$, $zx$) bands is less than 5\% as being irrelevant to the trend of $\bm S (\mu)$. 
This makes a clear contrast to the cubic SrTiO$_{3}$ case, where the three $t_{2g}$ bands have the equal contribution to $\bm S(\mu)$.\cite{kuroki.nacoo2.jpsj}

As already discussed by Kuroki {\em et al.}, the magnitude of Seebeck coefficient is mainly determined by the energy dependence of the electron velocity $\bm v_{n\bm k}$.\cite{kuroki.srtio3} 
By making a rough approximation, $\bm K_{0}$ and $\bm K_{1}$ in Eq.(2) are described as 
\begin{equation}
{\bm K_{0}} \approx \sum{ ({\bm v_{\rm high}}^{2}+{\bm v_{\rm low}}^{2})}, \ {\bm K_{1}} \approx k_{\rm B}T\sum{ ({\bm v_{\rm high}}^{2}-{\bm v_{\rm low}}^{2})}, 
\end{equation}
where the summation is over the states in the range of $| \varepsilon_{\bm k} - \mu | \lesssim k_{\rm B}T$ 
and ${\bm v_{\rm high}}$ and ${\bm v_{\rm low}}$ are typical velocities for the states above and below $\mu$, respectively.
Therefore, ${\bm v_{\rm high}} \gg  {\bm v_{\rm low}}$ situation  leads to a large Seebeck coefficients. 

Figure \ref{fig:fermi:v} (a) and (b) show the calculated  ${\bm v}_{n {\bm k}}^{2}$ superimposed on the sections of iso-energy surfaces $\varepsilon_{n{\bm k}}$ with several $\mu$ values. 
As increasing $\mu$, the ellipsoidal energy isosurface is elongated along the $k_{z}$ direction for $\mu <$ 0.2 eV,  and then the lobes grow up along the $k_{x}$ and $k_{y}$ direction for  $\mu >$ 0.2 eV. By taking a close look at Fig. \ref{fig:fermi:v} (a) and (b), the ellipsoidal energy surface has some amount of ${v}_{x}^{2}$ but almost zero ${v}_{z}^{2}$.  When $\mu >$ 0.2 eV, the energy surface starts to have finite ${v}_{z}^{2}$ value.  Therefore, the expectation value of  $\langle v_{z}^{2} \rangle (\mu)$ is shifted rightward by 0.2 eV with respect to  $\langle v_{x}^{2} \rangle (\mu)$  as shown in Fig.  \ref{fig:fermi:v} (c).  For $\mu \approx$ 0.2 eV, the situation ${\bm v_{z}^{ \rm high}} \gg  {\bm v_{z}^{ \rm low}}$  is realized so as to significantly increase $S_{zz}$. This can explain the ``bump'' behavior of $S_{zz} (x) $ in Fig. \ref{fig:seebeck:T} (b) 
at $x$=0.03 (the relation between $\mu$ and $x$ is shown in  Fig.  \ref{fig:fermi:v} (d)).

The ``bump'' behavior of $S(x)$ has been recently reported with the experimental observation in SnTe.\cite{zhou.snte.seebeckbump} 
SnTe is now known as a topological {\it crystalline} insulator, 
where the Sn-$p$ and Te-$p$ orbital characters are {\it inverted} between the valence-band top and conduction-band bottom at the R point.\cite{hsieh.snte.topological, paolo.snte.topological}  
When a hole carrier is doped, this  band inversion causes the warping in the Fermi surfaces of the valence band, which in turn leads to the enhancement of $S(x)$.\cite{singh.snte.seebeck, singh.gete} 
In short, in SnTe and the related system, the strong spin-orbit coupling mixing the orbital character between narrow gap causes non-harmonic Fermi surfaces, which is responsible for the high thermoelectric property. 
This is somehow related to the present case of BaTiO$_{3}$, where the polar crystal distortion asymmetries the Fermi surfaces and results in the sizable enhancement of $S(x)$ along the polar direction. 
By using this unique property, it may be possible to control the thermoelectric performance by the piezo-electric effect. 
When one imposes uniaxial strain on BaTiO$_{3}$, both the polar axis and the thermoelectrically favored direction may be manipulated as one wishes. 
This effect needs to be confirmed by further theoretical and experimental studies. 

\section{Conclusion}

In summary, within the Boltzmann transport theory  we calculated an anisotropic Seebeck coefficient  in ferroelectric BaTiO$_{3}$. 
The polar displacements of Ti and O ions cause the $t_{2g}$ band splitting into the ($yz, zx$) and the $xy$ states:  
the latter band in turn results in the strong anisotropy  and enhancement of the Seebeck coefficients. 
Such a structural distortion in perovskite oxides can provide a way to asymmetrize  the constant energy surfaces and 
the unusual energy dependence of the band velocity. 
This finding may pave a way for exploring novel thermoelectrics in various ferroelectric oxides. 
For example, PbTiO$_{3}$ may cause a larger Seebeck coefficient as having the huge electric polarization. 


\begin{acknowledgment}

This work is supported by a Grant-in-Aid for Young Scientists (B) from the Japan Society for the Promotion of Science (No 26800186) 
and 
by JST, CREST ``Creation of Innovative Functions of Intelligent Materials on the Basis of the Element Strategy''. 
The computation in this work has been done using the facilities of the
Supercomputer Center, Institute for Solid State Physics, University of Tokyo. 


\end{acknowledgment}



\end{document}